\documentclass[conference]{IEEEtran}
\IEEEoverridecommandlockouts
\usepackage{cite}
\usepackage{amsmath,amssymb,amsfonts}
\usepackage{algorithmic}
\usepackage{graphicx}
\usepackage{textcomp}
\usepackage{xcolor}
\usepackage{subcaption}
\def\BibTeX{{\rm B\kern-.05em{\sc i\kern-.025em b}\kern-.08em
    T\kern-.1667em\lower.7ex\hbox{E}\kern-.125emX}}
\begin{document}

\title{Distributed 3D Gaussian Splatting for High-Resolution Isosurface Visualization}

\author{\IEEEauthorblockN{1\textsuperscript{st} Mengjiao Han}
\IEEEauthorblockA{\textit{Argonne National Laboratory} \\
Lemont, IL, USA \\
hanm@anl.gov}
\and
\IEEEauthorblockN{2\textsuperscript{nd} Andres Sewell}
\IEEEauthorblockA{\textit{Utah State University} \\
Logan, UT, USA \\
a02024444@usu.edu}
\and
\IEEEauthorblockN{3\textsuperscript{rd} Joseph Insley}
\IEEEauthorblockA{\textit{Argonne National Laboratory} \\
Lemont, IL, USA \\
insley@anl.gov}
\and
\IEEEauthorblockN{4\textsuperscript{th} Janet Knowles}
\IEEEauthorblockA{\textit{Argonne National Laboratory} \\
Lemont, IL, USA\\
jknowles@anl.gov}
\and
\IEEEauthorblockN{5\textsuperscript{th} Victor A. Mateevitsi}
\IEEEauthorblockA{\textit{Argonne National Laboratory} \\
\textit{University of Illinois Chicago}\\
Lemont, IL, USA \\
vmateevitsi@anl.gov}
\and
\IEEEauthorblockN{6\textsuperscript{th} Michael E. Papka}
\IEEEauthorblockA{\textit{Argonne National Laboratory} \\
\textit{University of Illinois Chicago}\\
Lemont, IL, USA \\
papka@anl.gov}

\and
\IEEEauthorblockN{7\textsuperscript{th} Steve Petruzza}
\IEEEauthorblockA{\textit{Utah State University} \\
Logan, UT, USA \\
steve.petruzza@usu.edu}

\and
\IEEEauthorblockN{8\textsuperscript{th} Silvio Rizzi}
\IEEEauthorblockA{\textit{Argonne National Laboratory} \\
Lemont, IL, USA \\
srizzi@anl.gov}
}
\maketitle

\begin{abstract}
  3D Gaussian Splatting (3D-GS) has recently emerged as a powerful technique for real-time, photorealistic rendering by optimizing anisotropic Gaussian primitives from view-dependent images. While 3D-GS has been extended to scientific visualization, prior work remains limited to single-GPU settings, restricting scalability for large datasets on high-performance computing (HPC) systems.
    We present a distributed 3D-GS pipeline tailored for HPC. Our approach partitions data across nodes, trains Gaussian splats in parallel using multi-nodes and multi-GPUs, and merges splats for global rendering. To eliminate artifacts, we add ghost cells at partition boundaries and apply background masks to remove irrelevant pixels. Benchmarks on the Richtmyer–Meshkov datasets (about 106.7M Gaussians) show up to 3X speedup across 8 nodes on Polaris while preserving image quality. These results demonstrate that distributed 3D-GS enables scalable visualization of large-scale scientific data and provide a foundation for future in situ applications.
\end{abstract}


\section{Introduction}
3D Gaussian Splatting (3D-GS)~\cite{kerbl3Dgaussians} is a recent technique that enables photorealistic, real-time rendering of complex 3D scenes by optimizing anisotropic Gaussian primitives from view-dependent images. Unlike neural implicit methods such as NeRF, 3D-GS avoids the need for a neural network forward pass at inference, making it significantly faster while maintaining high visual fidelity.
Recent studies have applied 3D-GS to scientific visualization~\cite{ai2025nli4volvis,tang2025ivr,sewell2024high,yao2025volseggs}, but pipelines remain single-GPU, limiting scalability for large datasets. These datasets often exceed the memory of one GPU and are distributed across compute nodes on HPC, making centralized training impractical.
We address this gap with a distributed 3D-GS pipeline that supports multi-node, multi-GPU execution. Each node processes its local data subset, trains splats independently, and results are merged for global rendering. Additionally, we incorporate ghost cells and background masking to address rendering artifacts such as gaps and white streaks. The source code can be found at https://github.com/MengjiaoH/Grendel-GS-SciVIS. 

Our contributions are as follows.
\begin{itemize}
    \item \textbf{Distributed 3D-GS for HPC}: A multi-node, multi-GPU pipeline for large-scale scientific visualization, designed for datasets partitioned across HPC nodes. 
    \item \textbf{Scalability Benchmarks}: Performance evaluation across multiple datasets, image resolutions, and node counts.
    \item \textbf{Foundation for In Situ Visualization}: A first step toward scalable in situ workflows using 3D-GS for scientific data visualization.  
\end{itemize}

\section{Method}
\begin{figure}[!htbp]
    \centering
    \includegraphics[width=\linewidth]{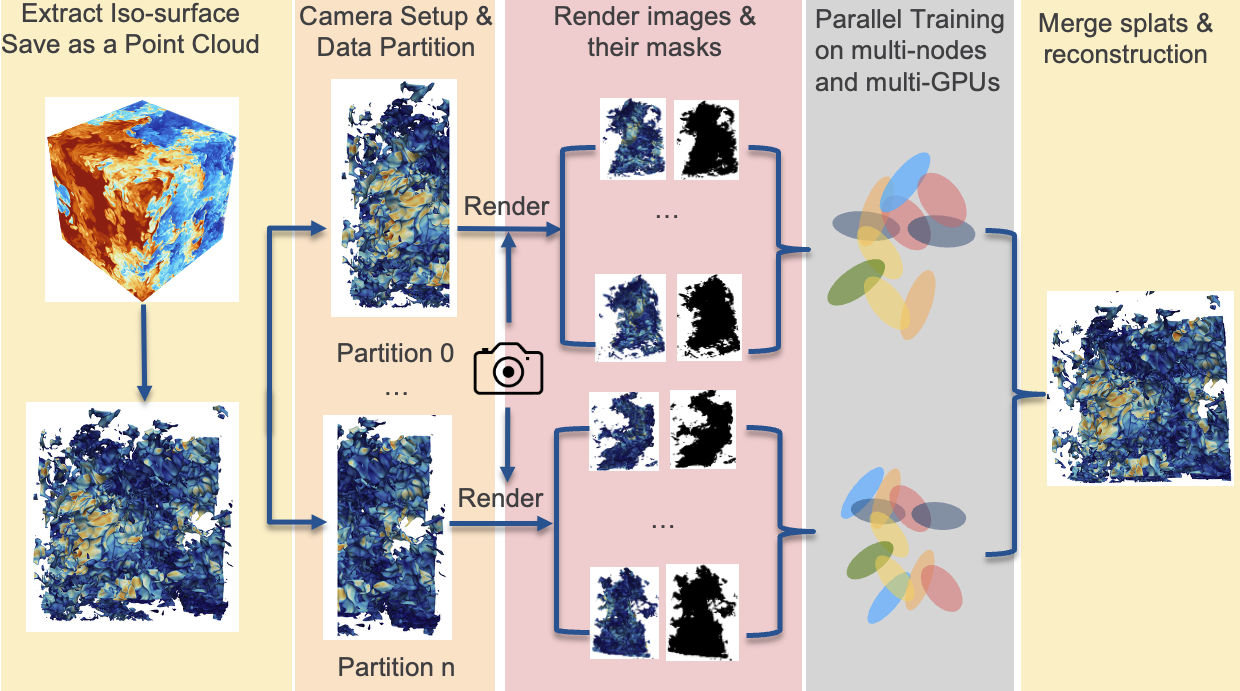}
    \caption{Workflow of our distributed 3D-GS pipeline for large-scale isosurface visualization on HPC systems.}
    \label{fig:pipline}
\end{figure}

Our distributed workflow (Figure \ref{fig:pipline}) proceeds as follows:
\begin{itemize}
    \item \textbf{Isosurface Extraction}: Use ParaView\footnote{https://www.paraview.org} to extract isosurface point clouds from volume datasets as initial Gaussian primitives. 
    \item \textbf{Camera Setup}: Generate a structured orbital set of synthetic camera views. All nodes use identical settings for training consistency.
    \item \textbf{Data Partitioning}: The dataset is divided into  $n$ partitions, one for each compute node, add ghost cells to avoid rendering gaps at partition boundaries.
    \item \textbf{Image Rendering and Masking}: On each node, render images and background masks for its own data partition. These masks prevent irrelevant splats and white streaks.
    \item \textbf{Parallel Training}: Each node trains a Gaussian splatting model independently using multi-GPU training approach~\cite{zhao2024scaling}.
    \item \textbf{Global Reconstruction}: Splats from all nodes are merged for final rendering.
\end{itemize}

This design removes the need to centralize large datasets, making distributed training tractable on HPC systems.

\section{Results}
 
We evaluated on three datasets: Kingsnake (110.3 MB, about 4M points)\footnote{https://www.digimorph.org/index.phtml}, Rayleigh–Taylor instability (491 MB, about 18.2M points)~\cite{miranda} and Richtmyer-Meshkov instability (5.3 GB, about 106.7M points)~\cite{richtmyer_meshkov}.
Training was performed on Polaris at Argonne\footnote{https://www.alcf.anl.gov/polaris}, with 4 NVIDIA A100 GPUs per node and 448 training images per dataset.

\subsection{Rendering Improvements}

\begin{figure}[!htbp]
    \centering
    \begin{subfigure}[t]{0.30\linewidth}
        \includegraphics[width=\linewidth]{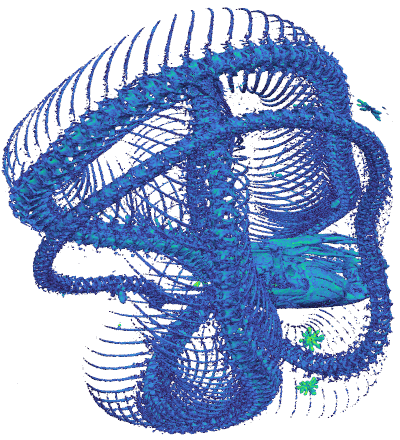}
        \caption{\centering Ground Truth}\label{img:a}
    \end{subfigure}
   \begin{subfigure}[t]{0.3\linewidth}
        \includegraphics[width=\linewidth]{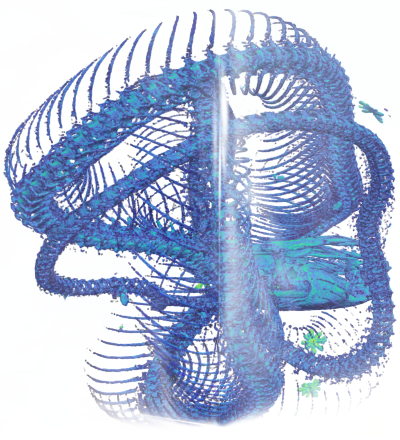}
        \caption{\centering W/O GC and Masks}\label{img:b}
    \end{subfigure}
    \begin{subfigure}[t]{0.3\linewidth}
        \includegraphics[width=\linewidth]{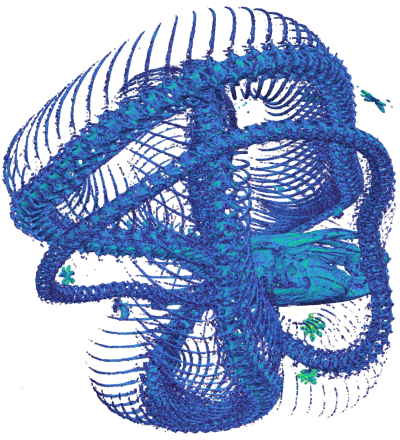}
        \caption{\centering Our Method W/ GC and Masks}\label{img:c}
    \end{subfigure}
    \caption{Visualization comparison using the Kingsnake scan dataset. (a) Ground truth, (b) rendering without ghost cells (GC) or background masks, and (c) our method with GC and background masks.
}
    \label{fig:improvements}
\end{figure}

Without our adjustments, merging splats across nodes introduced gaps and white streak artifacts (Figure \ref{img:b}). Adding ghost cells ensured smooth partition boundaries, while background masks removed unnecessary background splats, yielding artifact-free results (Figure~\ref{img:c}).

\subsection{Scaling Efficiency}
\subsubsection{Single Node Scaling}
\begin{figure}[!htbp]
    \centering
    \begin{subfigure}[t]{0.46\linewidth}
        \includegraphics[width=\linewidth]{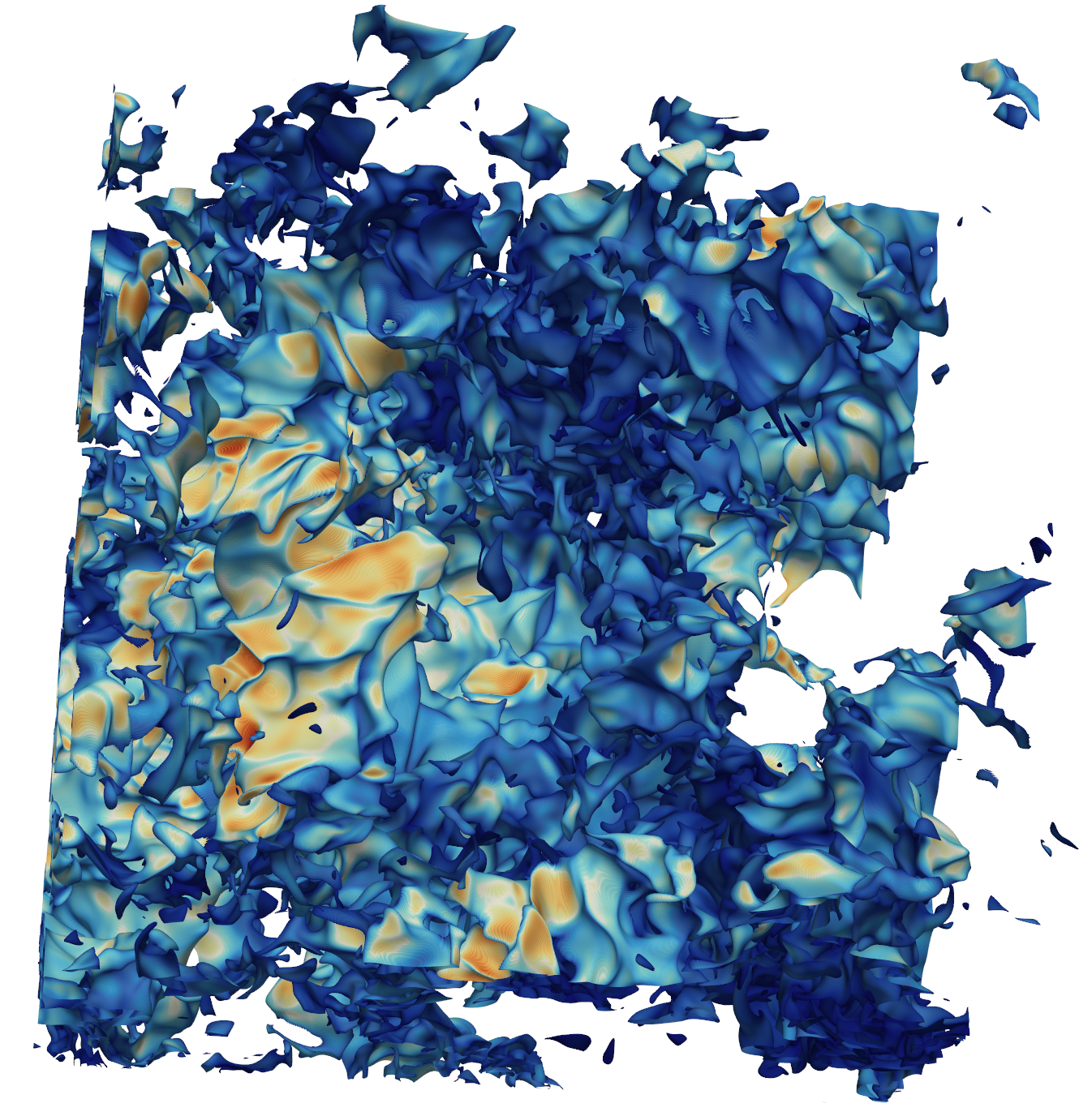}
        \caption{\centering Ground Truth}\label{img:a}
    \end{subfigure}
   \begin{subfigure}[t]{0.47\linewidth}
        \includegraphics[width=\linewidth]{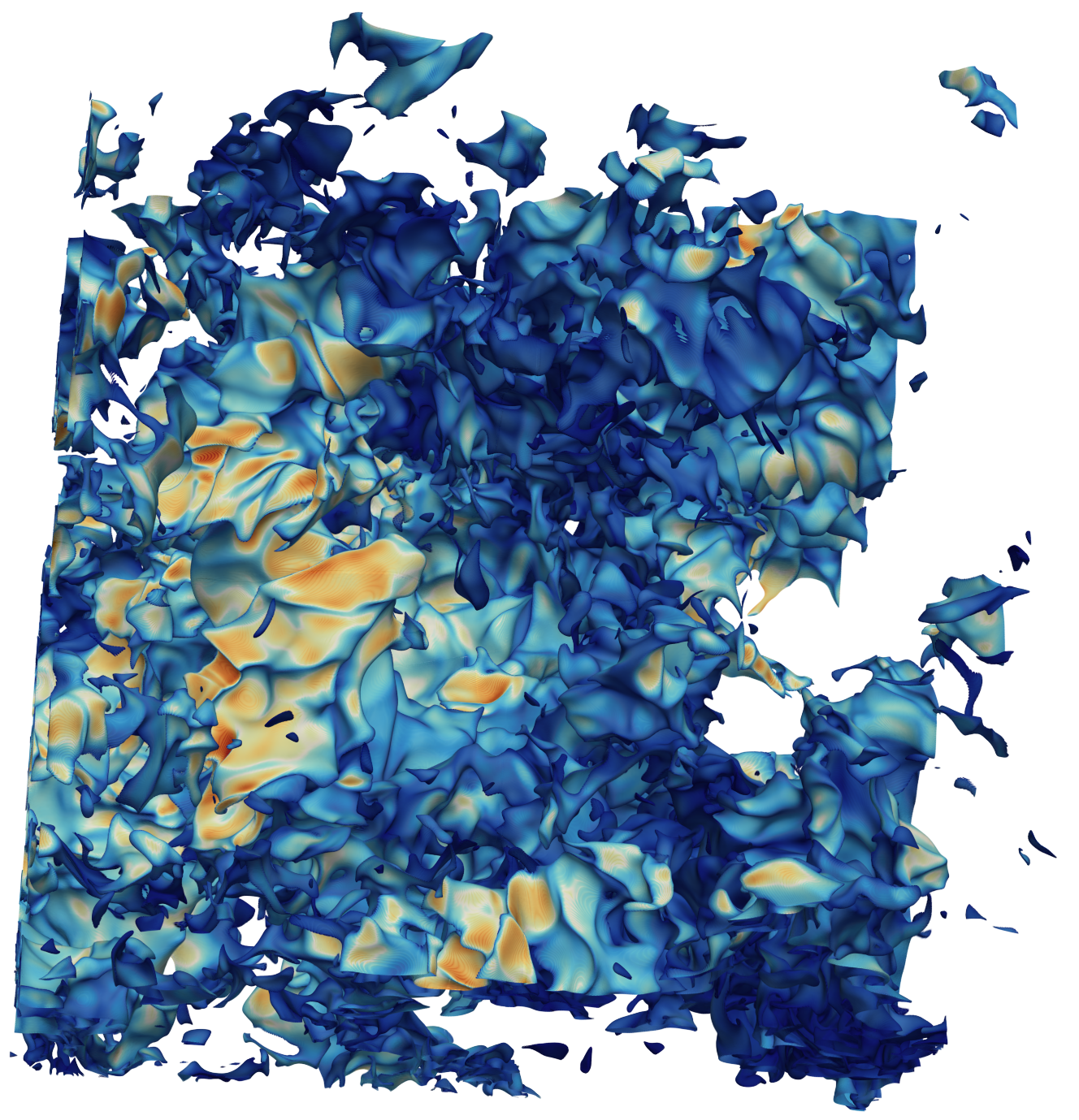}
        \caption{\centering Distributed 3D-GS}\label{img:b}
    \end{subfigure}
    \caption{Visualization on the Rayleigh–Taylor dataset: (a) ground truth and (b) reconstruction with our distributed 3D-GS method, achieving high fidelity (PSNR = 36.37, SSIM = 0.9905, LPIPS = 0.011).
}
    \label{fig:miranda-vis}
\end{figure}

On smaller datasets such as Kingsnake and Rayleigh–Taylor, multi-GPUs reduced training time significantly (Table \ref{tab:training_time}). For example, at $2048^2$ resolution, Kingsnake achieved a 5.6X speedup with 4 GPUs vs. 1, confirming effective intra-node scaling.
Since one A100 GPU can handle only 11.2M Gaussians~\cite{zhao2024scaling}, multi-GPU training was essential for larger datasets like Rayleigh–Taylor (Table \ref{tab:training_time}). 
Visualizations confirm high reconstruction quality (Figure~\ref{fig:miranda-vis}), with detailed metrics in Tables~\ref{tab:kingsnake} and \ref{tab:miranda}.

\begin{table}[]
\centering
\caption{Training time (minutes) for Kingsnake and Rayleigh–Taylor at different resolutions and GPU counts. ‘X’ marks runs that exceeded a single A100 GPU’s memory.}
\label{tab:training_time}
\resizebox{\columnwidth}{!}{%

\begin{tabular}{ccccc}
\hline
  & \multicolumn{2}{c}{Kingsnake ($\sim$4M)} & \multicolumn{2}{c}{Rayleigh–Taylor ($\sim$18M)} \\ \hline
\#GPUs & 1024 & 2048 & 1024 & 2048 \\ \hline
1           & 18.60           & 48.00             & X              & X              \\
2          & 10.48           & 15.46              & 21.88          & 50.10          \\
4          & 5.97            & 8.50               & 12.20          & 16.84          \\ \hline
\end{tabular}%
}
\end{table}

\begin{table}[h]
\caption{PSNR ($\uparrow$), SSIM ($\uparrow$), and LPIPS ($\downarrow$) for the Kingsnake dataset across different image resolutions and GPU counts.\vspace{-1em}}
\label{tab:kingsnake}
\resizebox{\columnwidth}{!}{%
\begin{tabular}{cccccccccc}
\hline
       & \multicolumn{3}{c}{512} & \multicolumn{3}{c}{1024} & \multicolumn{3}{c}{2048} \\ \hline
\#GPUs & PSNR   & SSIM  & LPIPS  & PSNR    & SSIM  & LPIPS  & PSNR    & SSIM  & LPIPS  \\ \hline
1      & 25.52  & 0.95  & 0.056  & 26.90   & 0.96  & 0.056  & 25.12   & 0.93  & 0.089  \\
2      & 25.87  & 0.96  & 0.046  & 28.63   & 0.97  & 0.035  & 29.33   & 0.97  & 0.030  \\
4      & 25.87  & 0.96  & 0.046  & 25.03   & 0.93  & 0.067  & 29.32   & 0.97  & 0.030  \\ \hline
\end{tabular}%
}
\vspace{-1em}
\end{table}

\begin{table}[]
\caption{PSNR ($\uparrow$), SSIM ($\uparrow$), and LPIPS ($\downarrow$) for the Rayleigh–Taylor dataset across different image resolutions and GPU counts.}
\label{tab:miranda}
\resizebox{\columnwidth}{!}{%
\begin{tabular}{cccccccccc}
\hline
       & \multicolumn{3}{c}{512} & \multicolumn{3}{c}{1024} & \multicolumn{3}{c}{2048} \\ \hline
\#GPUs & PSNR   & SSIM  & LPIPS  & PSNR    & SSIM  & LPIPS  & PSNR    & SSIM  & LPIPS  \\ \hline
2      & 31.62  & 0.99  & 0.014  & 34.21   & 0.99  & 0.010  & 36.30   & 0.99  & 0.011  \\
4      & 31.63  & 0.99  & 0.014  & 34.22   & 0.99  & 0.010  & 36.37   & 0.99  & 0.011  \\ \hline
\end{tabular}%
}
\end{table}

\begin{table}[]
\caption{Training time (minutes) for Rayleigh–Taylor and Richtmyer–Meshkov at different resolutions and node counts. ‘X’ = run failed due to memory limits.}
\label{tab:training-multinodes}
\resizebox{\columnwidth}{!}{%
\begin{tabular}{ccccc}
\hline
 & \multicolumn{2}{c}{Rayleigh–Taylor ($\sim$18M)} & \multicolumn{2}{c}{Richtmyer-Meshkov ($\sim$106.7M)} \\ \hline
\#Nodes & 1024 & 2048  & 1024  & 2048 \\ \hline
2       & 7.22 & 11.97 & X     & X    \\
4       & 5.08 & 8.33  & 10.07 & 32.03   \\
8       & 5.30 & 7.45    & 7.87  & 10.18   \\ \hline
\end{tabular}%
}
\end{table}

\begin{table}[]
\caption{PSNR ($\uparrow$), SSIM ($\uparrow$), and LPIPS ($\downarrow$) for the Rayleigh–Taylor dataset across different image resolutions and compute node counts.}
\label{tab:miranda-multinodes}
\resizebox{\columnwidth}{!}{%
\begin{tabular}{ccccccc}
\hline
        & \multicolumn{3}{c}{1024x1024}     & \multicolumn{3}{c}{2048x2048}     \\ \hline
\#Nodes & PSNR  & SSIM & LPIPS & PSNR  & SSIM & LPIPS \\ \hline
2      & 33.64 & 0.99 & 0.012 & 37.22 & 0.99 & 0.008 \\
4       & 33.61 & 0.99 & 0.012 & 37.15 & 0.99 & 0.008 \\ 
8       & 33.64 & 0.99 & 0.012 & 35.42 & 0.99 & 0.012 \\ \hline
\end{tabular}%
}
\end{table}

\subsubsection{Multi-Node Scaling}
\begin{figure}

    \begin{subfigure}[t]{0.48\linewidth}
    \centering
        \includegraphics[width=\linewidth, height=5cm, keepaspectratio]{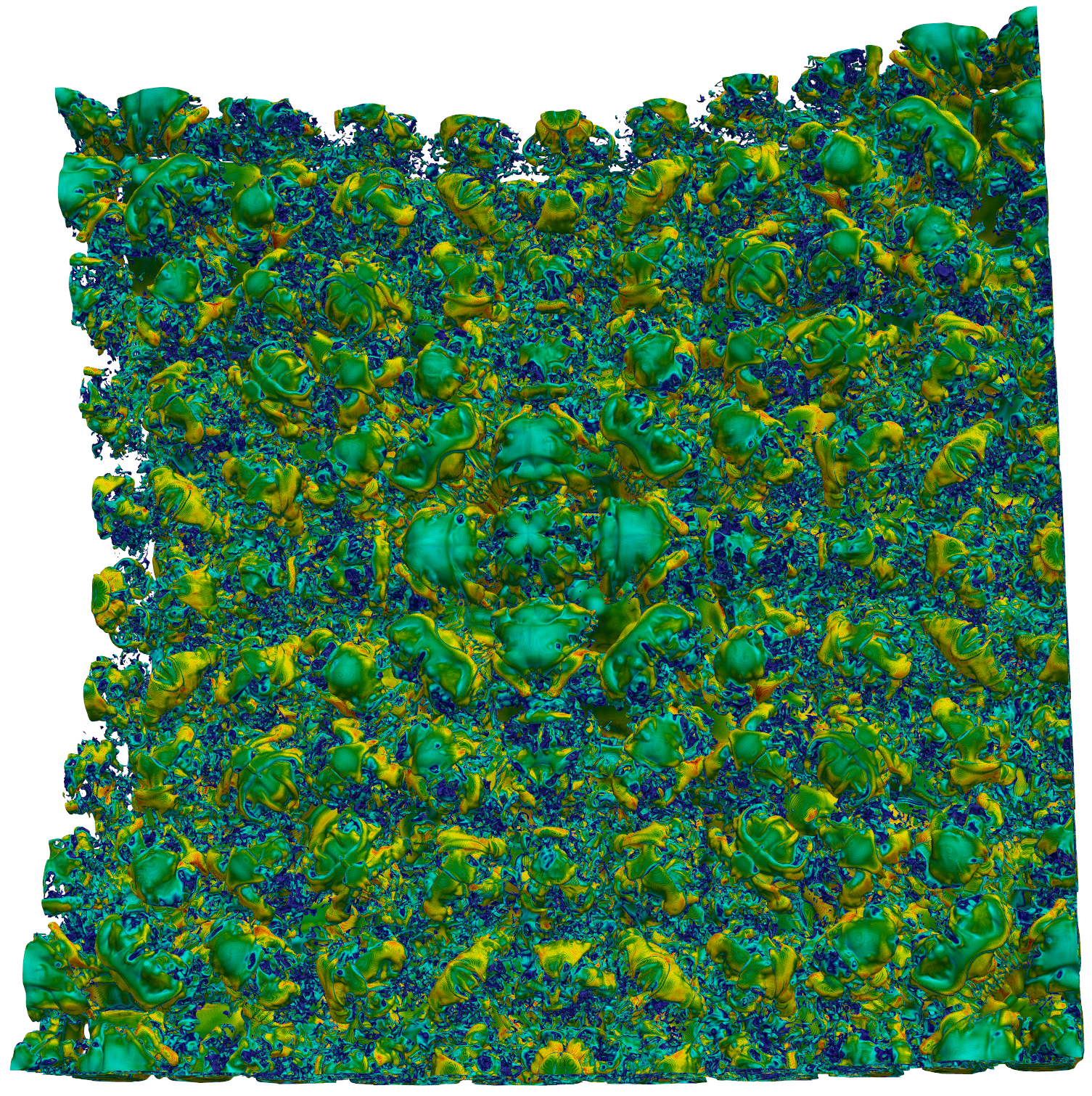}
        \caption{\centering Ground Truth}\label{img:a}
    \end{subfigure}
   \begin{subfigure}[t]{0.48\linewidth}
        \centering
        \includegraphics[width=\linewidth, height=5cm, keepaspectratio]{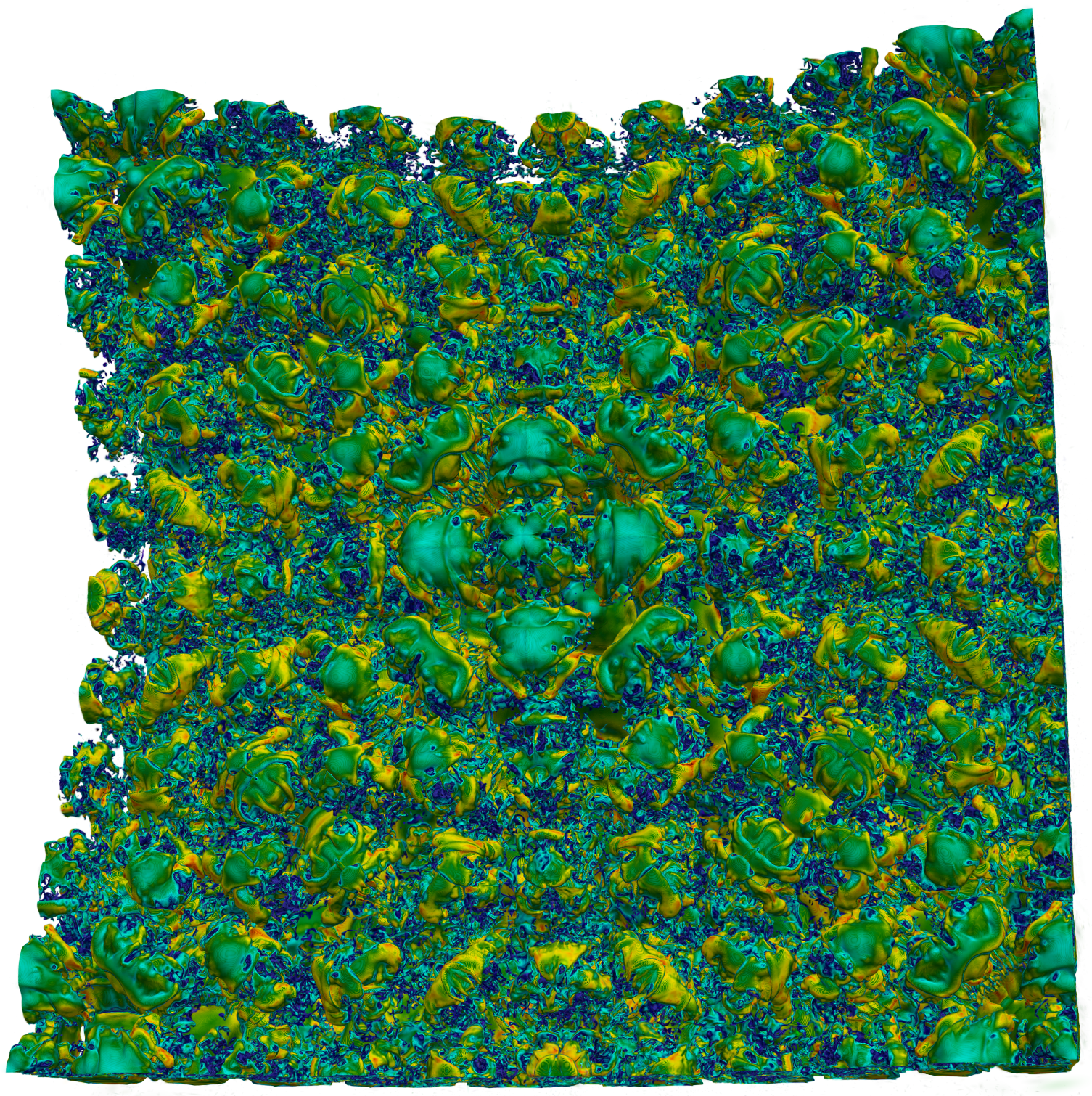}
        \caption{\centering Distributed 3D-GS Result}\label{img:b}
    \end{subfigure}
    \caption{Visualization of the Richtmyer–Meshkov instability dataset~\cite{richtmyer_meshkov} containing 106.7M Gaussians. (a) Ground truth image rendered directly from the point cloud. (b) Reconstruction from our distributed 3D-GS method at 2048×2048 resolution. Training was performed on 8 Polaris nodes at Argonne, each with 4 NVIDIA A100 GPUs, completing in 8 minutes. The reconstruction achieves high quality: PSNR=30, SSIM=0.97, LPIPS=0.019.
    }
  \label{fig:teaser}
\end{figure}

To evaluate scalability, we benchmarked our distributed 3D-GS pipeline on the Rayleigh–Taylor and Richtmyer–Meshkov datasets using multi-node runs with 4 GPUs per node.
For Rayleigh–Taylor, training scaled efficiently with a 1.4× speedup from 2 to 4 nodes (Table~\ref{tab:training-multinodes}), while maintaining high quality (Table~\ref{tab:miranda-multinodes}). However, because the dataset is not large enough, scaling from 4 to 8 nodes provided only limited gains in training time. 
The larger Richtmyer–Meshkov dataset required $\ge$ 4 nodes due to memory limits, but scaling to 8 nodes achieved a 3.1X speedup at $2048^2$ (Table~\ref{tab:training-multinodes}), with stable image quality (Table~\ref{tab:rm-multinodes}) and visualizations nearly identical to ground truth (Figure~\ref{fig:teaser}).
Overall, these results confirm that distributed 3D-GS scales effectively across multiple nodes, providing significant speedups while preserving visualization fidelity for both moderate and large-scale datasets.
%

\begin{table}[!t]
\caption{PSNR ($\uparrow$), SSIM ($\uparrow$), and LPIPS ($\downarrow$) for the Richtmyer-Meshkov dataset across different image resolutions and compute node counts.}
\label{tab:rm-multinodes}
\resizebox{\columnwidth}{!}{%
\begin{tabular}{ccccccc}
\hline
        & \multicolumn{3}{c}{1024x1024}     & \multicolumn{3}{c}{2048x2048}    \\ \hline
\#Nodes & PSNR  & SSIM & LPIPS & PSNR & SSIM & LPIPS \\ \hline
4       & 28.46 & 0.97 & 0.018 & 30.04   & 0.97   & 0.019    \\
8       & 28.20 & 0.96 & 0.019 & 30.04   & 0.97   & 0.019    \\ \hline
\end{tabular}%
}
\end{table}

\section{Conclusion and Future Work}
We introduced a distributed 3D-GS pipeline for large-scale scientific visualization on HPC. By partitioning datasets across nodes and incorporating ghost cells and background masks, we eliminate rendering artifacts while enabling multi-node, multi-GPU scaling. Our results show up to 3× faster training with consistently high quality.
Future work will extend this framework to real large-scale datasets, explore load balancing strategies, integrate with simulation pipelines for in situ rendering, and develop uncertainty quantification to provide scientists with confidence measures in reconstructed visualizations.

\section*{Acknowledgment}

This research used resources of the Argonne Leadership Computing Facility, a U.S. Department of Energy (DOE) Office of Science user facility at Argonne National Laboratory and is based on research supported by the U.S. DOE Office of Science-Advanced Scientific Computing Research Program, under Contract No. DE-AC02-06CH11357.

\bibliographystyle{IEEEtran}
\bibliography{conference_101719}

\end{document}